\documentclass[conference]{IEEEtran}

\usepackage{cite}
\usepackage{amsmath,amssymb,amsfonts}
\usepackage{graphicx}
\usepackage{xcolor}
\usepackage{booktabs}
\usepackage{array}
\usepackage{url}
\usepackage{algorithm}
\usepackage{algpseudocode}

\makeatletter
\def\@IEEEsectpunct{\ \,}
\makeatother

\makeatletter
\long\def\@makecaption#1#2{%
  \ifx\@captype\@IEEEtablestring%
  \footnotesize\bgroup\par\centering\@IEEEtabletopskipstrut{\normalfont\footnotesize #1}\\{\normalfont\footnotesize #2}\par\addvspace{0.5\baselineskip}\egroup%
  \@IEEEtablecaptionsepspace
  \else
  \@IEEEfigurecaptionsepspace
  \setbox\@tempboxa\hbox{\normalfont\footnotesize {#1.}\nobreakspace\nobreakspace #2}%
  \ifdim \wd\@tempboxa >\hsize%
  \setbox\@tempboxa\hbox{\normalfont\footnotesize {#1.}\nobreakspace\nobreakspace}%
  \parbox[t]{\hsize}{\normalfont\footnotesize\noindent\unhbox\@tempboxa#2}%
  \else%
  \ifCLASSOPTIONconference \hbox to\hsize{\normalfont\footnotesize\hfil\box\@tempboxa\hfil}%
  \else \hbox to\hsize{\normalfont\footnotesize\box\@tempboxa\hfil}%
  \fi\fi\fi}
\makeatother

\begin{document}

\raggedbottom

\title{Evidence-Guided Repository-Level RTL Repair}

\author{%
\IEEEauthorblockN{Yuxin Du\textsuperscript{1}, Juxin Niu\textsuperscript{1},
Zhe Jiang\textsuperscript{2,3}, and Nan Guan\textsuperscript{1}}%
\IEEEauthorblockA{\textsuperscript{1}City University of Hong Kong, Hong Kong, China\\
\textsuperscript{2}Southeast University, China\\
\textsuperscript{3}National Center of Technology Innovation for EDA, China\\
Email: yuxindu8-c@my.cityu.edu.hk, juxin.niu@my.cityu.edu.hk,
101013615@seu.edu.cn, nanguan@cityu.edu.hk}%
}

\maketitle

\begin{abstract}
Repository-level RTL repair must localize a failure that spans files, modules,
and clock cycles, then propagate the fix consistently. Existing methods reason over source code, which reveals
possible behaviors but not the failed execution, and cannot tell whether a local
fix was propagated consistently. We therefore present an evidence-guided
framework with three modules. Failure grounding converts a problem statement into a
reproduced failing run and a failure anchor. Waveform-guided localization uses a
localization toolbox to narrow the observed violation into a candidate
mechanism and an evidence trail. Consistency-aware repair and validation then
expand that seed into a coordinated patch and replay the same scenario to check
that the violation disappears. We conducted experiments on HWE-Bench and achieved better performance than the baseline.
\end{abstract}



\section{Introduction}

Debugging RTL designs consumes much of the effort in chip
development~\cite{foster2015trends,gogotb2026}. Doing it
by hand is tedious~\cite{lik2024}. An engineer must trace the bug across files and modules, and
every hypothesis needs another test run~\cite{meic2024,veripilot2026}. Agents suit this
task and are already applied to RTL debugging with promising
results~\cite{verilogcoder,fvdebug,veritrace,bluesfl,llm4sechw2024}. That progress, however, is
confined to the module level~\cite{cirfix,rtl-repair,veribug2024,veribugbench2026}. At that level, the design unit and its testbench
are supplied. Repository-level RTL bug repair instead takes a repository and a
problem statement. The output is a patch that makes the repository pass the
fail-to-pass tests. HWE-Bench~\cite{hwe-bench} evaluates general-purpose agents
on this setting, but no published method targets it.

RTL debuggers typically trace an observed failure back to its cause and then
fix it~\cite{dekleer1987,reiter1987,feldman2014,rodler2019}. This recipe gets much harder at repository scale. This task differs from
small-scale repair in two ways. First, a repository contains many files and
modules. The agent therefore needs a long reasoning chain to reach the root
cause. That chain also crosses clock cycles. No specification or reference
design is available to confirm a candidate cause~\cite{chockler2004,halpern2005causes}.
Correct localization
becomes difficult~\cite{granularity2026,headroom2026}. Second, repository-level bugs usually span several modules
and files. Such a bug is solved only when every affected place is changed
consistently. That requirement makes repair hard as well.

\begin{figure}[t]
  \centering
  \includegraphics[width=\columnwidth]{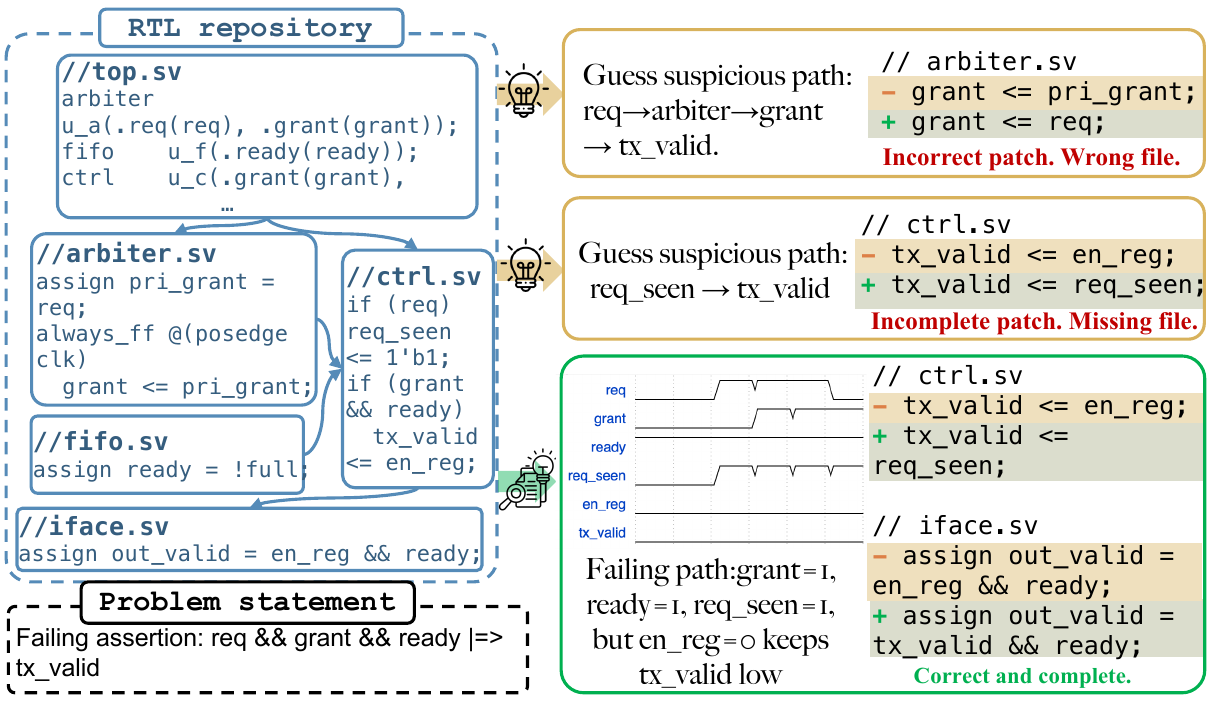}
  \caption{Without cycle-level evidence, the agent tests hypotheses that are
  consistent with everything it can observe.}
  \label{fig:motivation}
\end{figure}

These two difficulties have a common cause: the agent lacks grounded evidence.
Current debugging methods lean heavily on reasoning over source code. For
small-scale RTL debugging, that ability is already strong enough. At repository
scale it stops working. Source code shows every path the design could take, but
not the one this failing run took~\cite{goldmine2010,hertz2013assertions,assertionbench}. Hardware correctness is decided as the state
evolves cycle by cycle, so many candidate causes fit everything the agent can
read from the repository~\cite{waveformqa}. The agent can only test one at a time, and the loop
does not converge. Figure~\ref{fig:motivation} shows this from a real
trajectory. Even after a local cause is found, the agent still lacks a reliable
way to tell whether the same failing scenario has been preserved while the
violation itself disappears. A patch can remove the observed symptom and still
leave stale interfaces, widths, or state updates elsewhere. We therefore shift
the approach from reasoning over source code alone to evidence-guided bug
repair. The key evidence is the reproduced failing run and its waveform.

We present a framework for this task. It organizes the method into three
modules. Failure grounding turns the problem statement into a reproduced failing run and a
failure anchor that records the scenario, the observed violation, and the run to
replay. Waveform-guided localization then uses a localization toolbox to bound the
violation in time, trace active drivers, and lift raw values into design-level
semantics, yielding a candidate mechanism plus an evidence trail.
Consistency-aware repair and validation finally expand that local seed into a coordinated
repository patch and replay the same scenario to check that the violation
disappears without breaking the surrounding behavior. Because the framework
reuses one observed execution throughout, it grounds both diagnosis and repair
checking in what the design actually did rather than in source-only reasoning.

We evaluate this framework on HWE-Bench, which poses real problem statements from large
RTL repositories and judges each patch by fail-to-pass tests. It collects 417
tasks; 29 OpenTitan cases cannot be adapted to a runnable VCS environment in our
setup, so every condition runs over the same 388 tasks. The framework resolves
more of them than the baseline for every model we test.

\section{Related Work}

\subsection{Automated RTL Repair}

Traditional hardware repair tools such as CirFix~\cite{cirfix} and
RTL-Repair~\cite{rtl-repair} repair a given module against a supplied failing
testbench, and other work debugs inside generation or verification at that scale.
R3A pairs multi-agent fault localization with a search that keeps only
verified patches~\cite{r3a2025}, Clover dispatches repair operations across
agents and symbolic solvers under the same search~\cite{clover2026}, and
MechMem-RTL reuses verifier-confirmed repair records whose stored trigger
matches the new failure~\cite{mechmem2026}. VerilogCoder adds AST-based waveform
tracing to graph planning against golden testbenches~\cite{verilogcoder}, VeriTrace
queries signals and time windows~\cite{veritrace}, and FVDebug works from formal
counterexamples and specifications~\cite{fvdebug}. BluesFL instead stops at
localization, reading internal signals along co-simulation-guided slices of a
19K-line design~\cite{bluesfl}.
HWE-Bench moves to the repository level, giving an agent a problem statement and the whole
repository and judging its fix by project-native flows, so it contributes a benchmark
rather than a method~\cite{hwe-bench}. Our work targets this setting, where the faulty
module and a self-contained reproduction are not given, and repairs may cross file
and module boundaries.

\subsection{Repository-Level Coding Agents}

Repository-level coding systems retrieve, call tools, and validate to address issue
reports, and SWE-agent designs the interface that carries them~\cite{swe-agent}.
Agentless fixes a pipeline of localization, repair, and validation~\cite{agentless};
graph-guided agents explore adaptively, LocAgent over a heterogeneous entity
graph~\cite{locagent}, GraphLocator over causal links from sub-issues to code
entities~\cite{graphlocator}, and repository memory by adding commits, issues, and
summaries~\cite{repomemory}. These systems target software code repositories; hardware
repositories instead rely on knowledge graphs over Verilog sources and their
descriptions~\cite{verbkg}. Our work moves this setting to end-to-end repair of
existing RTL repositories, where reproduction, localization, and validation must be
established in project-specific flows.


\section{Methodology}

\subsection{Overview}
\label{sec:overview}

\begin{figure*}[!t]
  \centering
  \includegraphics[width=\textwidth]{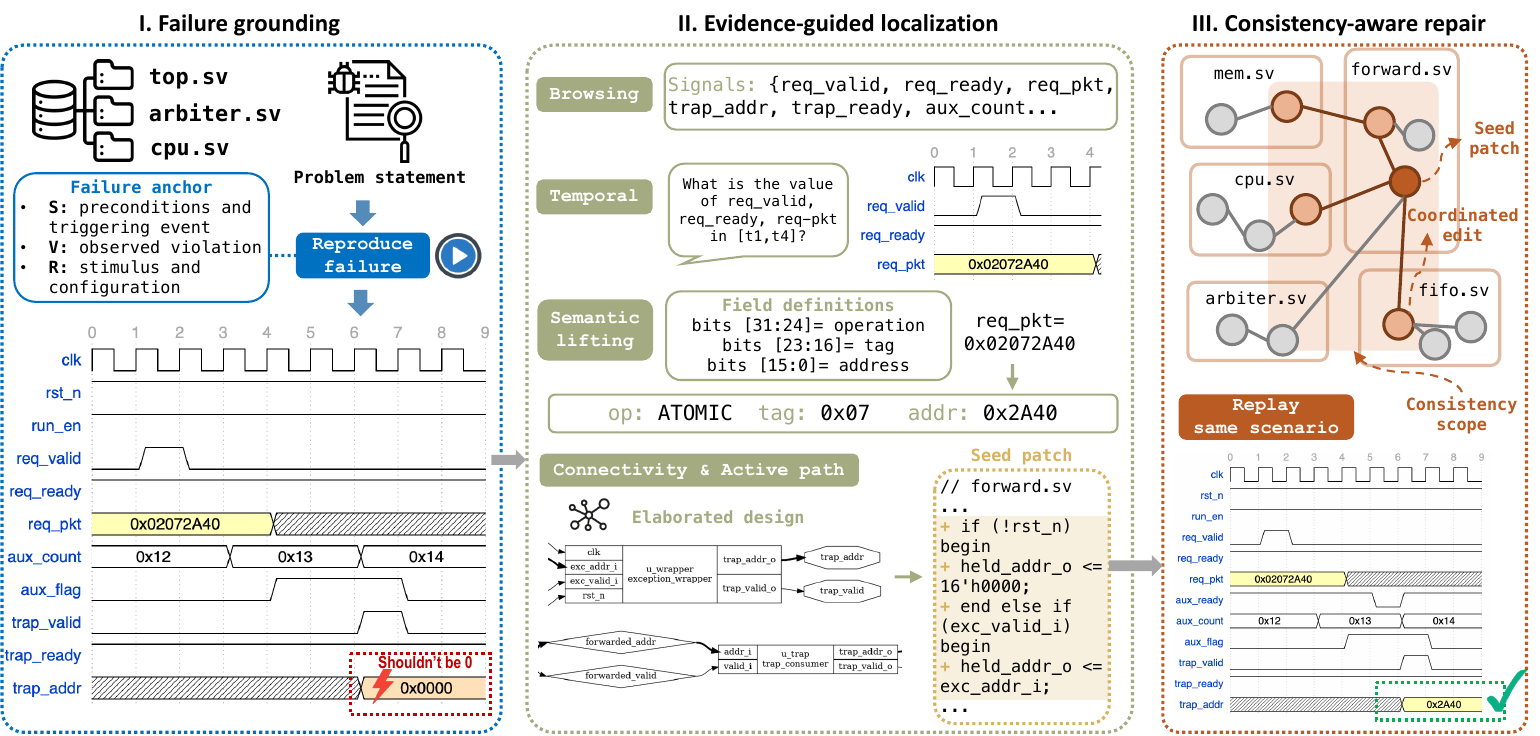}
  \caption{The framework runs one evidence-guided loop. It reconstructs a failing run and dumps its
  waveform, helps the agent localize a candidate mechanism with the localization toolbox, and then
  checks whether a coordinated patch preserves the scenario while removing the observed violation.}
  \label{fig:overview}
\end{figure*}

Figure~\ref{fig:overview} shows the evidence-guided loop. The method centers on one shared
artifact, the waveform of a reproduced failing run. Failure grounding
first turns the problem statement into a concrete failing execution and its waveform.
Waveform-guided localization then gives the agent a reusable localization toolbox for extracting, tracing, and
semantically interpreting signal behavior from that execution. Consistency-aware repair and
validation finally turn a local diagnosis into coordinated repository edits and check whether the
revised design removes the same violation while preserving the reproduced scenario.
Section~\ref{sec:repro} details failure grounding, Section~\ref{sec:localization} presents
waveform-guided localization, and Section~\ref{sec:repair} describes consistency-aware repair and
validation.

\subsection{Failure Grounding}
\label{sec:repro}

\paragraph{Failure specification extraction.}
The held-out fail-to-pass tests are not visible to the agent, so the framework must first construct its
own failing run. To keep that run tied to the problem statement, the framework parses it into three checkable
elements: the preconditions $P$ under which the bug should appear, the triggering event or
transaction $E$, and the required behavior $Q$ that is violated.

\paragraph{Grounded reproduction.}
A reproduced run is useful only if later localization reasons about the same failure mechanism
described in the problem statement. The framework therefore constructs one failing execution for the reported
scenario and enables tracing on the unmodified repository. The resulting run passes an
alignment check and is accepted as grounding evidence only when the waveform shows that $P$ and $E$
actually occur and that $Q$ is observably violated. The framework also requires the stimulus and configuration to stay within the repository's
intended usage assumptions, such as legal parameter settings, interface protocols, and supported
operating modes. Unclear assumptions are recorded explicitly instead of being promoted to facts.

\paragraph{Failure anchor formulation.}
Later modules need one stable representation of the reproduced failure; otherwise localization and
validation may drift to different scenarios. Failure grounding therefore outputs a failure anchor
$F=(S,V,R)$ with three elements:
\begin{itemize}
  \item $S=(P,E)$ is the reproduced situation, where $P$ collects the preconditions under which the
  bug should appear and $E$ denotes the triggering event or transaction.
  \item $V=(Q,\ell,t)$ is the observed violation, where $Q$ is the required behavior, contradicted at
  location $\ell$ and cycle or interval $t$.
  \item $R$ stores the stimulus and configuration used to reproduce the failure.
\end{itemize}
Localization later consumes $(V,R)$ to explain the observed symptom, while validation replays $R$
and checks that $S$ still occurs after a candidate repair and that $V$ disappears. The anchor keeps
later diagnosis and validation aligned to the same reproduced failure.

\subsection{Waveform-Guided Localization}
\label{sec:localization}

Given the failure anchor $F=(S,V,R)$, localization takes the observed violation $V$ in reproduced
run $R$ as input and returns $L=(M,T)$, where $M$ is a candidate mechanism and $T$ is an evidence
trail. The challenge is that the waveform is large while only a small part is causally relevant. The
framework therefore provides a reusable localization toolbox that makes the failing execution
systematically
queryable together with the elaborated design under the reproduced configuration. This setup
supports queries across module boundaries, instantiations, and generated hierarchy.

\paragraph{Waveform query interface.}
Raw waveforms are too large and heterogeneous for direct free-form inspection, so localization
needs a small set of structured queries that progressively narrow the search space. Localization
therefore uses five query classes, summarized in Table~\ref{tab:query_primitives}. Together, they
move the analysis from a symptom signal, to a bounded interval, to the exercised logic, and finally
to design-level semantics.

\begin{table}[t]
  \centering
  \caption{Query classes the localization toolbox provides, their role, and their
  returned artifacts.}
  \label{tab:query_primitives}
  \footnotesize
  \setlength{\tabcolsep}{3pt}
  \renewcommand{\arraystretch}{0.97}
  \begin{tabular}{p{0.175\columnwidth}p{0.37\columnwidth}p{0.33\columnwidth}}
    \toprule
    Query class & Localization role & Artifact \\
    \midrule
    Browsing & Align problem-statement or source names with waveform hierarchy. & Scope/signal candidates. \\
    Temporal & Bound the violation window. & Divergence windows and timelines. \\
    Connectivity & Trace from symptom signals to producing logic. & Driver/load/fan-in links. \\
    Active path & Recover the executed dependency chain. & Executed driver chain. \\
    Semantic lifting & Map raw values to RTL-level semantics. & Decoded fields, symbolic states, and aligned waveforms. \\
    \bottomrule
  \end{tabular}
\end{table}


Beyond structural queries, localization also relies on semantic lifting to turn raw waveform
values into design-level concepts. This includes decoding packed fields, mapping state encodings to
symbolic states, and aligning related waveforms on a shared timeline. These lifted artifacts let the
agent reason about candidate mechanisms at the RTL semantic level rather than at the level of raw
integers.

\paragraph{Evidence-narrowing loop.}
Even with a query interface, localization remains ambiguous unless evidence is used to eliminate
competing explanations step by step. Localization therefore applies these primitives as a loop: it
time-bounds the violation in $V$, traces backward over active drivers, invokes semantic lifting
when raw values are not yet interpretable, and widens the waveform window or compares alternatives when the
current evidence is insufficient. It terminates only when it can summarize a candidate
mechanism $M$ together with an evidence trail $T$ that records both supporting observations and
remaining ambiguities. Because each narrowing step is backed by values observed in $R$, the final
mechanism remains linked to an explicit evidence trail.

\subsection{Consistency-Aware Repair and Validation}
\label{sec:repair}

Consistency-aware repair addresses the second repository-level failure mode: a correct local diagnosis
does not yet imply a repository-consistent fix. Once one site changes, the same mechanism may
still be represented in connected definitions, bindings, or replicated realizations, so the fix
must be propagated and rechecked under the same reproduced scenario. Given $F$ and localization
output $L=(M,T)$, the framework builds a coordinated patch $C$, expands it across structurally linked
locations, and validates that the original scenario is preserved while the reported violation is
removed.

\paragraph{Repository relation graph.}
To determine where a seed repair must propagate, the framework constructs a repository relation graph
over repair sites. Nodes denote repair sites at the declaration, implementation, structural-binding,
and generated-RTL levels, together with their enclosing modules or files when needed for
propagation. Edges encode structural relations that may require coordinated updates across these
sites.

The graph combines source-level and elaboration-level evidence. Source parsing recovers symbol
relations, structural containment, and local dependency information. HDL elaboration recovers the
resolved hierarchy and concrete binding contexts, including instantiated connectivity, parameter
resolution, width resolution, and generated replicas, thereby turning source templates into
concrete repair contexts. For generated RTL, the framework links emitted RTL locations back to the
generator source or
shared template that produced them, using source maps, line directives, or generator metadata when
available. When such information is unavailable, the framework adds conservative structurally matched
links and marks them as uncertain. Starting from the sites edited by the seed patch, closure over
$G_R$ yields the candidate propagation scope.

\begin{table}[t]
  \centering
  \caption{Instantiation sites the source declares against the instances the
  configured design contains, for five modules taken from the cases we
  elaborated.}
  \label{tab:elaboration}
  \small
  \setlength{\tabcolsep}{6pt}
  \begin{tabular}{llrr}
    \toprule
    Repository & Module & source & elaborated \\
    \midrule
    Ibex      & ibex\_csr                  &  21 &  15 \\
    Ibex      & ibex\_counter              &   3 &  12 \\
    CVA6      & lzc                        &  39 & 115 \\
    CVA6      & spill\_register\_flushable &   1 &  60 \\
    OpenTitan & prim\_subreg               & 104 & 104 \\
    \bottomrule
  \end{tabular}
\end{table}

Table~\ref{tab:elaboration} shows why the relation graph follows the elaborated
design: counting textual instantiation sites overstates the scope when a site is
never built, as for \emph{ibex\_csr} behind a disabled PMP gate, and understates
it when one site expands into many, as for CVA6's
\emph{spill\_register\_flushable}. A closure over source-level relations therefore
demands edits at places this configuration never instantiates and misses
replicated instances it does, so the scope it returns is not the scope the design
has, and a seed repair propagated over it is incomplete in exactly the way
repository-level bugs are. For generated RTL the graph also links emitted RTL
back to its generator, and OpenTitan's counts are a lower bound: no
Verilator model can be built for it, so the graph is built from a partial pyslang netlist without
parameter values.

\begin{algorithm}[t]
  \caption{Consistency-aware repair and validation}
  \label{alg:repair}
  \footnotesize
  \begin{algorithmic}[1]
    \Require $F=(S,V,R)$, $L=(M,T)$, $G_R$
    \Ensure validated patch $C$
    \State $(P,E) \gets S$ \Comment{$P$: reproduced preconditions; $E$: trigger}
    \State $(Q,\ell,t) \gets V$ \Comment{$Q$: required behavior}
    \State $C \gets \Call{SeedPatch}{L}$
    \Repeat
      \Statex \textbf{Phase I: Consistency-scope construction}
      \State $E_C \gets \Call{EditedSites}{C}$
      \State $U \gets \Call{Closure}{E_C,G_R}$
      \State $U_{\mathrm{edit}} \gets \Call{SelectRequired}{U,C,T}$
      \State $C \gets \Call{PropagateRepair}{C,U_{\mathrm{edit}}}$
      \Statex \textbf{Phase II: Replay-based validation}
      \State $O \gets \Call{Replay}{R,C}$
      \If{$\neg \Call{PreserveScenario}{O,P,E}$}
        \State $C \gets \Call{RefinePatch}{C,O,L}$
      \ElsIf{$\neg \Call{EliminateViolation}{O,Q}$}
        \State $U \gets \Call{ExpandScope}{U,G_R,O}$
        \State $U_{\mathrm{edit}} \gets \Call{SelectRequired}{U,C,T}$
        \State $C \gets \Call{PropagateRepair}{C,U_{\mathrm{edit}}}$
      \EndIf
    \Until{$\Call{PreserveScenario}{O,P,E} \land \Call{EliminateViolation}{O,Q}$}
    \State \Return $C$
  \end{algorithmic}
\end{algorithm}

\paragraph{Consistency closure and validation.}
Algorithm~\ref{alg:repair} summarizes this loop in two phases. In Phase~I, the agent uses $G_R$
to enumerate the sites that are structurally coupled to its edited locations. This traversal
intentionally over-approximates the consistency scope, so the agent next inspects the reached
sites and marks the subset that truly requires synchronized changes. In Phase~II, the resulting
coordinated patch is validated by replaying the reproduction recipe $R$ from $F$. A patch is
accepted only if the replay still realizes the reproduced preconditions $P$ and trigger $E$, and
the required behavior $Q$ is no longer violated. If replay breaks $P$ or $E$, then the current
patch has changed the reproduced operating condition and the seed repair itself must be refined. If
$P$ and $E$ are preserved but $Q$ is still violated, then the local edit remains insufficient and
the consistency scope should be expanded before the next attempt.


\section{Experiments}
\label{sec:experiments}

\subsection{Experiment Setup}
\label{sec:setup}

\paragraph{Benchmark and metric.}
We evaluate on HWE-Bench~\cite{hwe-bench}, which collects 417 real bug-fixing
tasks from six open-source RTL repositories. Each task provides a repository at a
buggy revision and a problem statement, and success is judged by the repository's
own held-out fail-to-pass tests. We report the resolved rate, the share of tasks
whose held-out fail-to-pass tests all pass. The tasks cover RISC-V cores, SoCs, processors,
and security roots of trust. OpenTitan needs a commercial simulator (VCS) while
the other five repositories run under Verilator, and 29 of OpenTitan's 245 tasks
cannot be adapted to a runnable VCS environment in our setup. Those 29 are
skipped for every condition, so every result below is over the remaining 388
tasks, the agent never sees the held-out tests, and all denominators use the same
pool.

\paragraph{Implementation.}
For the baseline we run each model under its own scaffold, unchanged:
DeepSeek V4 Flash (0731) under OpenHands SDK, DeepSeek V4 Pro under Claude Code,
and Kimi K2.6 under Kimi CLI.

Relative to the baseline, the framework adds three modules. First, failure
grounding reproduces a failing run and stores it as an FST waveform. Second,
waveform-guided localization exposes a localization toolbox whose scripts browse
the waveform hierarchy, bound the violation in time, trace connectivity and the
active path, and lift raw values into RTL-level semantics. Third,
consistency-aware repair and validation propagate a local patch to structurally
coupled sites and then replay the reproduced scenario to check the result.

\subsection{Results and Analysis}
\label{sec:results}

We read the experiments around four questions and close with what the framework
costs to run.\\
\noindent\hspace*{1em}\textbf{RQ1:} Does the framework resolve more tasks?\\
\hspace*{1em}\textbf{RQ2:} How much does each arm contribute?\\
\hspace*{1em}\textbf{RQ3:} What changes in how the agent works?\\
\hspace*{1em}\textbf{RQ4:} Why do the remaining failures persist?

\subsubsection*{RQ1: Does the framework resolve more tasks?}
\label{sec:exp_rq1}

\begin{table*}[t]
  \centering
  \caption{Resolved tasks per repository; each cell is
  \emph{resolved}/\emph{tasks}.}
  \label{tab:main_results}
  \small
  \setlength{\tabcolsep}{4pt}
  \begin{tabular}{llccccccc}
    \toprule
    Model & Scaffold & OpenTitan & Ibex & CVA6 & Caliptra & XiangShan & Rocket Chip & Total \\
    \midrule
    \multicolumn{9}{l}{\emph{Baseline}} \\
    DeepSeek V4 Flash (0731) & OpenHands SDK & 148/216 & 28/35 & 32/35 & 16/16 & 29/54 & 20/32 & 273/388 \\
    DeepSeek V4 Pro          & Claude Code   & 132/216 & 20/35 & 30/35 & 10/16 & 32/54 & 19/32 & 243/388 \\
    Kimi K2.6                & Kimi CLI      & 130/216 & 28/35 & 30/35 & 13/16 & 34/54 & 23/32 & 258/388 \\
    \midrule
    \multicolumn{9}{l}{\emph{Evidence-guided (ours)}} \\
    DeepSeek V4 Flash (0731) & OpenHands SDK & 169/216 & 30/35 & 33/35 & 16/16 & 38/54 & 26/32 & 312/388 \\
    DeepSeek V4 Pro          & Claude Code   & 157/216 & 28/35 & 32/35 & 16/16 & 40/54 & 24/32 & 297/388 \\
    Kimi K2.6                & Kimi CLI      & 160/216 & 31/35 & 33/35 & 16/16 & 37/54 & 26/32 & 303/388 \\
    \bottomrule
  \end{tabular}
\end{table*}

Table~\ref{tab:main_results} shows the main result: every model improves on its
baseline, and the gain is spread over every repository where the baseline leaves
headroom, concentrating on the two largest. OpenTitan is a hardware security root
of trust whose deep module hierarchy and generated register layers spread one
behavioral change over many files; XiangShan is an out-of-order RISC-V core
written in Chisel, so a fix usually has to reach many parameterized instances.
Both carry the most headroom, and most of their missed tasks need edits in
several files, the coordinated setting the consistency module is built for.
Across the pool every model improves on its own scaffold, so the gain is the
framework's rather than a stronger agent's. These failures need a mechanism and a
complete patch, and the framework supplies both, from a reproduced run and the
elaborated scope.

\subsubsection*{RQ2: How much does each arm contribute?}
\label{sec:exp_rq2}

\begin{figure}[t]
  \centering
  \includegraphics[width=\columnwidth]{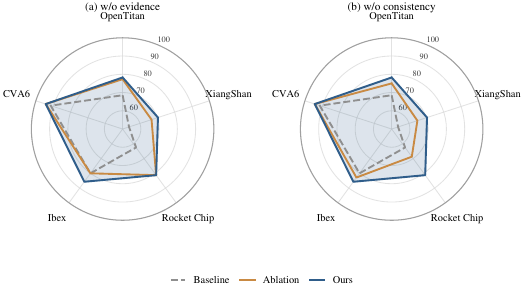}
  \caption{Resolved rate per repository, DeepSeek V4 Flash (0731), under the
  baseline, the framework, and the framework with one arm ablated. The
  radial axis is percent and starts at 50\%.}
  \label{fig:contribution}
\end{figure}

Figure~\ref{fig:contribution} reads the pool three ways per repository: the
baseline, the framework, and the framework with one arm ablated. The evidence
arm drops the alignment check and the localization toolbox; the
consistency arm drops scope construction and replay validation. Either ablation
costs the framework repairs, so neither arm is redundant. The evidence arm
works from a single recorded
execution, so it pays off only where the failure is visible in that waveform. The consistency arm reads the repository and the elaborated hierarchy, so it
catches the coupled edits no waveform shows and loses repairs on Rocket Chip,
where the evidence arm loses none. Between the two, the consistency arm
matters most: the repairs it recovers need edits at coupled modules, and
locating those modules is itself a form of localization that a single waveform cannot supply. The two arms
therefore cover different halves of the job, evidence supplying the mechanism and
consistency deciding whether the patch built on it reaches every module the fix
requires.

\subsubsection*{RQ3: What changes in how the agent works?}
\label{sec:exp_behaviour}

\begin{figure}[t]
  \centering
  \includegraphics[width=\columnwidth]{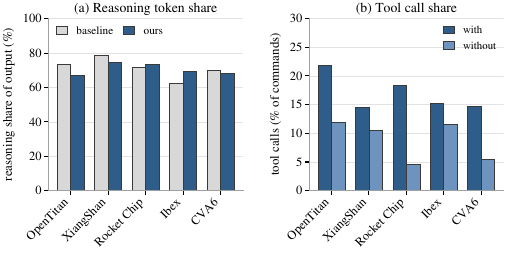}
  \caption{Agent behavior per repository with DeepSeek V4 Flash (0731):
  reasoning token share under the framework and the baseline (a), and tool call
  share with and without the localization toolbox (b).}
  \label{fig:tool_behaviour}
\end{figure}

\paragraph{Reasoning share.}
Reading a waveform does not displace the agent's reasoning.
Figure~\ref{fig:tool_behaviour}a compares the share of a run's output that is
reasoning text. The median share falls on OpenTitan and XiangShan, rises on
Rocket Chip and Ibex, and barely moves on CVA6; every shift stays inside the
spread the baseline shows across its own cases. With the toolbox in hand the agent
therefore stays a reasoner rather than an executor and combines reasoning with
execution, which is what turns the evidence into repairs.

\paragraph{Tool call share.}
Figure~\ref{fig:tool_behaviour}b reports the share of the agent's commands that
are tool calls, and with the toolbox mounted that share is higher than in the
arm without it, so the agent reaches for it on its own. The arm without it calls
tools too, but those are scripts
the agent writes itself, and nothing guarantees that they are correct. Either the agent
notices a broken script and rewrites it, which wastes tokens and puts queries too
complex to write by hand out of reach, or the script is syntactically sound but
semantically wrong, and the values it returns mislead the repair. Under the framework the toolbox scripts are verified, so a call returns
evidence the agent can act on and its tool use is efficient.

\subsubsection*{RQ4: Why do the remaining failures persist?}
\label{sec:exp_failures}
\setcounter{paragraph}{0}

\begin{table}[t]
  \centering
  \caption{Failures the DeepSeek V4 Flash (0731) baseline leaves unresolved,
  by the evidence available. The three columns partition a repository's
  failures and are given in percent.}
  \label{tab:grounding}
  \small
  \setlength{\tabcolsep}{5pt}
  \begin{tabular}{lrrr}
    \toprule
    Repository & accepted & rejected & none \\
    \midrule
    OpenTitan   & 58.8 &  4.4 & 36.8 \\
    XiangShan   & 56.0 & 24.0 & 20.0 \\
    Rocket Chip & 41.7 &  8.3 & 50.0 \\
    Ibex        & 57.1 & 14.3 & 28.6 \\
    CVA6        & 66.7 &  0.0 & 33.3 \\
    \midrule
    Total       & 56.5 &  9.6 & 33.9 \\
    \bottomrule
  \end{tabular}
\end{table}

Table~\ref{tab:grounding} splits the failures the baseline leaves unresolved per
repository by the evidence available for them: the share the alignment check
accepts, the share it rejects, and the share with no waveform at all. The
\emph{none} share is where the repository's own simulation flow cannot be driven
in our setup, and the \emph{rejected} share where a waveform exists but is too
sparse, or holds a level or subtree that does not carry the failure. They persist for two reasons.

\paragraph{Where the evidence does not help.}
These failures need information no run of the buggy revision carries, either in
the interface the task defines or in the RTL generator. On \emph{Ibex-104}
the held-out test compiles against three CSR ports that revision does not contain, so
the repair must invent an interface whose names live only in the test and in the
golden patch, the repository's own fix that the agent never sees. A waveform
cannot name a port that was never built. On
\emph{RocketChip-1493} the waveform shows the asynchronous crossing behind the stalled
boot-ROM access and the agent reroutes the boot ROM in one Scala file, while the
golden patch introduces a control bus and migrates the peripherals onto it, a
generator-level decision that no value in the waveform ranks above the smaller edit.

\paragraph{Where the agent does not use it well.}
Where a waveform existed, the agent usually consulted it and still failed. That step
needs the protocol or micro-architectural contract behind the signals, which the
agent must supply from hardware knowledge, and where that knowledge is incomplete
the repair follows the values to the wrong edit. On \emph{XiangShan-1242} the
queries return the comparison that stalls a load, and the agent reads the empty
store-set pointer as a missing assignment instead of seeing that the pointer is
defined on the ROB index, which the golden patch rewrites eleven Scala files to
carry. On \emph{OpenTitan-22878} it reassembles the key-manager payload in the file
the problem statement implicates but keeps the original field order and leaves the valid-side
and debug-invalid mappings alone, because the values show which fields are wrong
and nothing about the order and handshake the protocol fixes.

\subsubsection*{Cost.}
\label{sec:exp_limits}

\begin{figure}[t]
  \centering
  \includegraphics[width=\columnwidth]{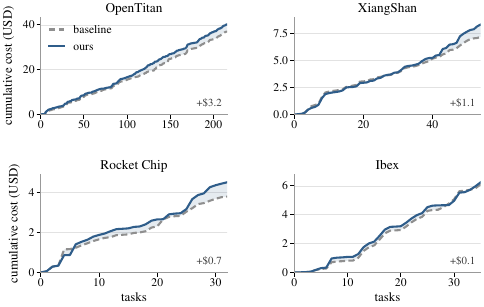}
  \caption{Cumulative model spend per repository with DeepSeek V4 Flash (0731):
  the baseline and the framework, with the shaded band showing the marginal
  spend the framework adds.}
  \label{fig:cost}
\end{figure}

Figure~\ref{fig:cost} draws the two curves per repository, and they
stay close on all four. The framework adds a thin band on OpenTitan and
XiangShan, less on Rocket Chip, and on Ibex the two are hard to tell apart, so no
repository pays much for the evidence. The waveform is written once and read
through scripts, which the agent never loads as a file, so a query returns the few
values it asks for and a waveform of tens of thousands of signals costs the
tokens of one short command. The toolbox also cuts what the baseline spends on
guessing, since the waveform lets the agent test a hypothesis instead of reading
files on suspicion, and those passes offset what the queries add. OpenTitan grows most and is the
most complex, with the largest waveforms.

\section{Conclusion}
Repository-level RTL repair must localize failures across files, modules, and clock cycles and then propagate fixes consistently, without a specification or reference design to certify either step. We presented an evidence-guided framework whose three modules ground the repair in a reproduced failing run, narrow the observed violation into a candidate mechanism and an evidence trail, and expand that seed into a coordinated, replay-checked patch. On the 388 HWE-Bench tasks we evaluate, the framework resolves more tasks than the baseline for every model we test. Trajectories show agents extracting signal values, lifting them into design semantics, and using reruns to expose incomplete repairs, while remaining failures highlight the gap between obtaining relevant evidence and producing complete repository-level fixes.

\clearpage
\bibliographystyle{IEEEtran}
\bibliography{references}




\end{document}